\documentclass[twocolumn,showpacs,amsmath,amssymb, prb]{revtex4}

\usepackage{graphicx}
\usepackage{dcolumn}
\usepackage{bm}

\begin{document}

\title{Magnetization dynamics in optically excited nanostructured nickel films}

\author{Georg M. M\"uller}
 \email{mueller@nano.uni-hannover.de}
 \altaffiliation[Current address: ]{Institut f\"ur Festk\"orperphysik, Leibniz Universit\"at Hannover, Appelstra\ss{}e 2, D-30169 Hanover, Germany}
\author{Gerrit Eilers}%
\author{Zhao Wang}
\author{Malte Scherff}
\affiliation{IV. Physikalisches Institut, Universit\"at G\"ottingen, Friedrich-Hund-Platz 1, D-37077 G\"ottingen, Germany}%

\author{Ran Ji}
\author{Kornelius Nielsch}
\altaffiliation[Current address: ]{Institut f\"ur Angewandte Physik, Universit\"at Hamburg, Jungiusstra\ss{}e 11, D-20355 Hamburg, Germany}
\affiliation{
Max-Planck-Institut f\"ur Mikrostrukturphysik, Weinberg 2, D-06120 Halle, Germany}

\author{Markus M\"unzenberg}
\affiliation{IV. Physikalisches Institut, Universit\"at G\"ottingen, Friedrich-Hund-Platz 1, D-37077 G\"ottingen, Germany}

\date{\today}

\begin{abstract}
In this work, Laser-induced magnetization dynamics of nanostructured nickel films is investigated. The influence of the nanosize is discussed considering the time-scale of hundreds of femtoseconds as well as the GHz regime. While no nanosize effect is observed on the short time-scale, the excited magnetic mode in the GHz regime can be identified by comparison with micromagnetic simulations. The thickness dependence reveals insight on the dipole interaction between single nickel structures. Also, transient reflectivity changes are discussed.
\end{abstract}

\pacs{75.75.+a, 75.30.Fv }
\maketitle
Nowadays, nanostructured ferromagnetic films are a key component of hard disks as well as magnetic random access memories.  Furthermore, new concepts in which such films are utilized for logical operations by means of the stray field interaction between single elements are on their way.\cite{Cowburn02252000} In all these applications, the speed of certain operations is limited by the time needed to switch the magnetization of a single element. With regard to this technological impact, the magnetization dynamics of nanosized ferromagnetic elements has become a major field of research.

In 1996, Beaurepaire et al. demonstrated the demagnetization within hundreds of femtoseconds of a nickel film as a response to a femtosecond laser pulse.\cite{Beaurepaire1996} Since then, not only the physical origin of this effect  has been discussed and examined extensively, but also this demagnetization mechanism has become a tool to induce and study magnetic motion of ferromagnetic films on a much longer time scale.\cite{Kampen2002}

\begin{figure}[h!]
		\includegraphics[width=0.8\columnwidth]{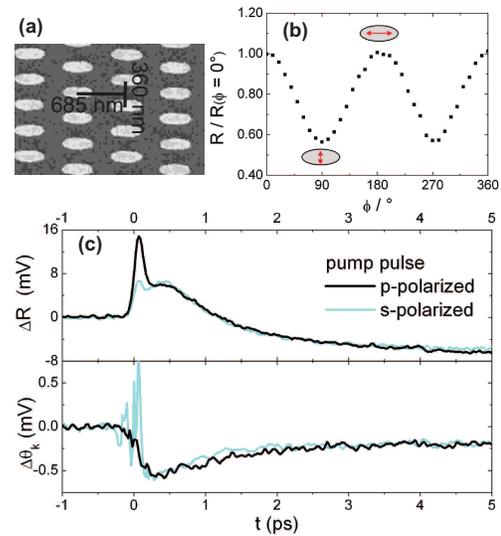}
	\caption{\label{fig1}(Color online) (a) SEM image of the elliptically nanostructured nickel film ; the long (short) axis averages to 320 (130) nm.(b) Reflectivity of the pump light as a function of its plane of polarization; p-polarized (parallel to the long axis) pump light has an increased reflectivity. (c) Transient reflectivity and Kerr angle changes for s- and p-polarized pump light; the enhancement of the reflectivity due to plasmonic resonance during pump pulse duration finds no counterpart in the Kerr transients.}
\end{figure}

In this work, the transient change of magnetization and reflectivity of nanostructured nickel films in response to an femtosecond pump pulse is examined. We investigate the dynamics on a time scale of hundreds of femtoseconds  as well as the transient changes in the GHz regime. Several reports on similar experiments have been lately published.\cite{comin:217201,Lepadatu2007, Kruglyak2007} Here, we vary structure size or pump light polarization to directly study nanosize effects. On the short time scale, we observe no influence of the nanoscale on the ultrafast demagnetization while the transient reflectivity is enhanced due to plasmonic resonance during pump pulse duration. Further, we compare the experimentally examined magnetization dynamics on the longer timescale of a nanostructured film with the dynamics of a single nanoelement found  by micromagnetic simulations. Especially, the thickness  dependence is studied which allows to quantitatively determine the influence of the dipole interaction between single nanodiscs.

The thin film nanodot arrays are prepared by laser interference lithography (IL). Details about the IL steps\cite{jmmm:316:e44}  and the resist stack\cite{Ji2006,Redondo2006}, which is applied here, can be found elsewhere. Electron beam evaporated Nickel is deposited in ultrahigh vacuum at a base pressure of $5\cdot 10^{-10}$ mbar  and capped by 2 nm of Magnesium oxide; as substrate thermally oxidized silicon is used. The remaining photoresist is removed by chemical etching in 1-methyl-2pyrrolidinone at 120$^{\circ}$C. In our experiment, probe and pump pulse are generated by a Titanium:sapphire fs laser together with a regenerative amplifier (repetition rate 250 kHz, pulse width of 60 - 80 fs, spot size of probe and pump pulse about $40$ and $60\,\mathrm{\mu m}$, respectively). For the time resolved Kerr measurements, we utilize a double modulation scheme so that the polarization of the probe pulse is modulated with a photo-elastic modulator (PEM) and the intensity of the pump pulse with a mechanical chopper.  The sample is mounted at room temperature with the external magnetic field is applied in the plane of incidence.

The influence of the nanoscale on the time-scale of femtoseconds is discussed with respect to the dynamics we observed in an  elliptically nanostructured nickel film. The nickel thickness amounts to 26 nm. The long (short) axis of the ellipses averages to about 320 (130) nm. A Scanning Electron Microscope (SEM) image of this film is shown in Fig. \ref{fig1} (a) where also the distances  between single elements are given.

Due to plasmonic resonance,\cite{wokaun1982} the reflectivity of the elliptic structures strongly depends on the polarization of the light. In Fig. \ref{fig1} (b), the reflectivity of the pump pulse as a function of its light polarization is plotted; the reflectivity is higher for p-polarized (parallel to the long axis of the ellipse) light than for s-polarized light. This antenna effect also becomes manifest on the transient change of reflectivity as it is depicted in Fig. \ref{fig1} (c): During pump pulse duration the transient reflectivity of the probe pulse is enhanced for both polarization states of the pump light, but by far more pronounced in the case of p-polarization. After pump pulse duration, the two curves look very similar. This nanosize effect does not become manifest in the transient change of the Kerr angle that is depicted in the bottom graph of Fig. \ref{fig1} (c). It is ensured that these curves contain no transient reflectivity changes since they are formed as the difference of two Kerr transients with the external field of $\pm 100$ mT  applied in opposite direction along the sample surface. The difference between the two demagnetization curves stems from a measurement artifact, the so called coherent artifact\cite{Eichler1986} that originates from pump probe interference and is much more pronounced in our setup for s-polarized light due to the probe light modulation with the PEM.

This observation could help to understand the direct influence of the photonic or plasmonic field on the ultrafast demagnetization. In an early model, Zhang and H\"ubner explained the ultrafast demagnetization by taking directly into account the laser field.\cite{PhysRevLett.85.3025} The enhanced polarization within the nanostructure for $p$-polarized light should directly become manifest in a higher demagnetization which is not the case. Therewith, our experimental observation sets limits how the laser field should be incorporated in a microscopic  model of Laser-induced demagnetization.\footnote{Koopmans et al. already discussed the photonic influence by means of conservation of angular momentum.\cite{Koopmans2000}}

\begin{figure}
		\includegraphics[width=0.9\columnwidth]{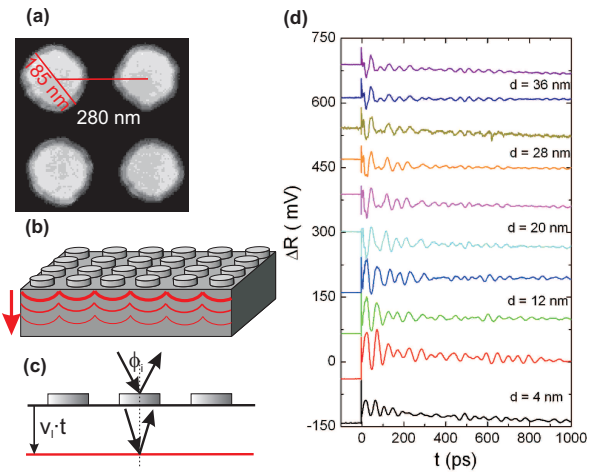}
	\caption{\label{fig2}(Color online) (a) SEM image of the examined structure. Heat transfer into the substrate leads to the formation of standing surface acoustic waves (b) and also to interference effects for reflected probe light (c). (d) Traced reflectivity changes for the different nickel thicknesses.}
\end{figure}

For the discussion of the dynamics in the GHz regime, the magnetization and reflectivity transients of a circularly structured nickel wedge are presented. This structure contains quadratically ordered nickel discs of a diameter of about 185 nm with an interdot distance of 280 nm. The slope of the nickel thickness is 5 nm/mm. A SEM image is given in Fig. \ref{fig2} (a).

\begin{figure}
		\includegraphics[width=0.9\columnwidth]{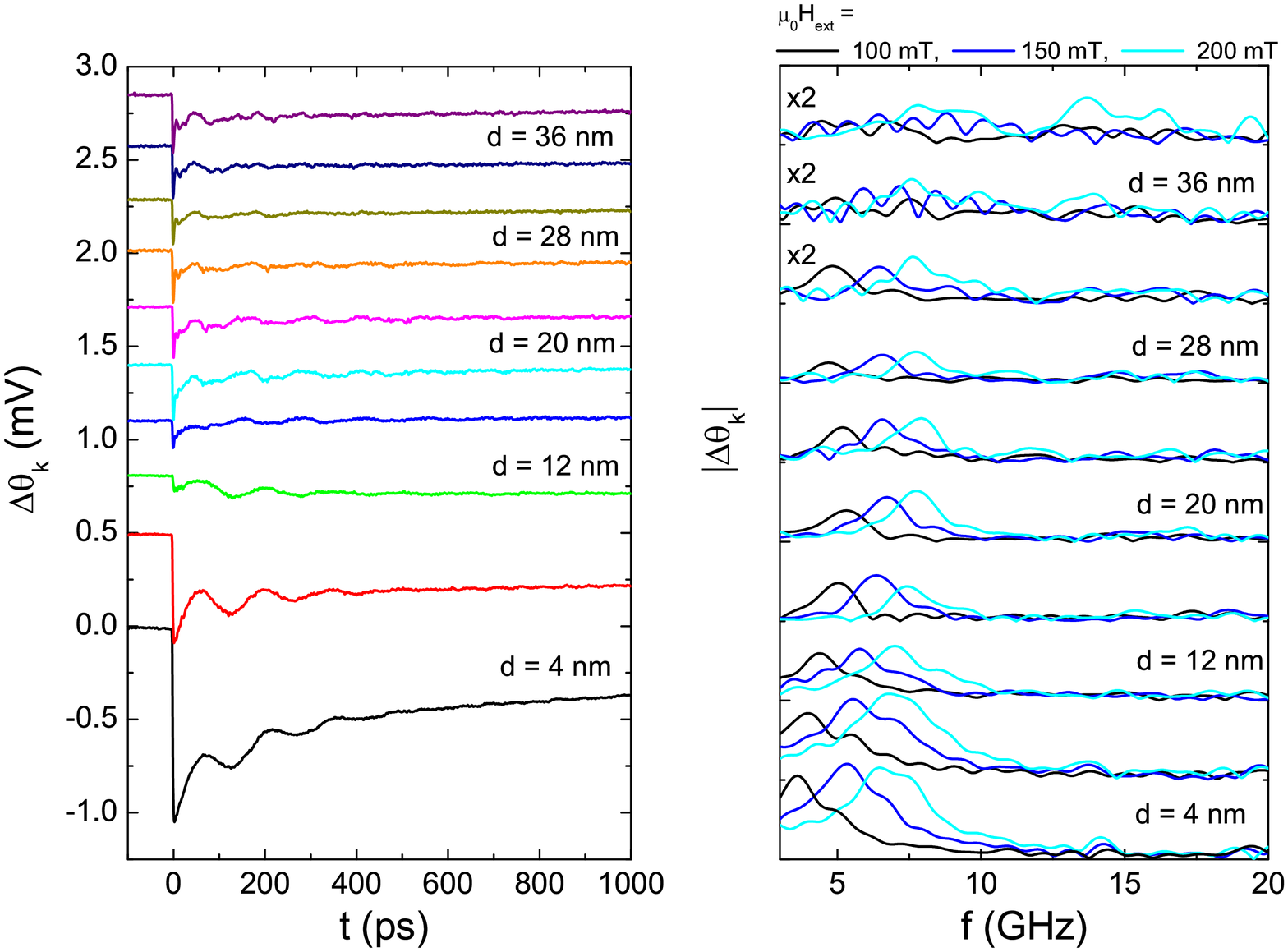}
	\caption{\label{fig3}(Color online) Time resolved change of the Kerr angle for the different nickel thicknesses at an external of 200 mT and correspondng Fourier transforms for 100, 150, and 200 mT.}
\end{figure}
The reflectivity transients are given in Fig. \ref{fig2} (d). Reflectivity changes of such nanostructures in optical pump probe experiments due to the heat transport from the surface into substrate have  already been studied since years opening the field of picosecond ultrasonics.\cite{lin:37} According to the work by Lin et al., we can attribute the oscillatory reflectivity changes to at least two different effects: (i) In the first hundred picoseconds, probe light reflected from the sample surface interferes with probe light that gets reflected from the wave front of the strain wave propagating into the substrate as it is schematically depicted in Fig.\ref{fig2} (d). According to Lin et al., the reflectivity oscillates due to this effect with a frequency of
\begin{equation}
f=\frac{2v_{\mathrm{l}}\sqrt{(\Re n)^2 -\sin^2\phi_{\mathrm{i}}}}{\lambda}
\end{equation}
 where $\phi_{\mathrm{i}}\approx 30^{\circ}$ denotes the angle of incidence and $n$ and $v_{\mathrm{l}}$ refractive index and longitudinal sound velocity of the substrate,  respectively. As the substrate is thermally oxidized silicon, this effect should blur out after some tens of picoseconds. Taking the material parameters of crystalline quartz $\Re n=2.53$ for 800 nm and $v_{\mathrm{l}}=5970\,\mathrm{m/S}$, the frequency calculates to 21.7 GHz, which corresponds roughly to the oscillation in the first hundred picoseconds in  the time domain. (ii) Furthermore, the strain wave propagating  into the substrate and the thermal expansion of the nanodiscs lead to enhanced stress below the metallic structures. Therewith, on the one hand, elastic modes will be excited within single metallic structures and, on the other hand,  a standing surface acoustic wave will be formed in the substrate due to the periodicity of the structures. According to the discussion in the cited reference,\citep{lin:37} this elastic mode can be viewed in two limiting cases either as a sole standing surface acoustic wave of the substrate or  as separate vibrations of single metallic discs. 
Recently, Giannetti et al. studied the coupling of these two systems in detail.\cite{giannetti:125413}  In our case, the frequencies of the oscillations in the beating patterns given in Fig. \ref{fig2} (d) that we attribute to this effect shift towards to lower values as the thickness increases. This is a clear indicator that these dynamics cannot be viewed in the limiting case of pure standing surface acoustic waves. Therefore, effects stemming from magneto-elastic coupling might be incorporated in the Kerr  dynamics for which  some indications were found in a work by Comin et al.  \cite{comin:217201} 
 
For the separation of the reflectivity changes from the Kerr dynamics, we have to trace two Kerr transients with external magnetic fields of opposite sign. Therewith, we only can detect magnetic modes that change their phase by $\pi$ under switching the external field which might not be valid for all modes that get excited in this complex system.\cite{muller:020412} For excitation of magnetic precessional motion, the external magnetic field is applied in an angle of 30$^{\circ}$ to the sample surface.\cite{Kampen2002} It is assured that in the used field range the magnetization of a nanodisc is given by a single domain state for all thicknesses.     The Kerr transients for the different thicknesses are  depicted in Fig. \ref{fig3} for an external field value of 200 mT. Also, Fourier transforms for field values of 100, 150, and 200 mT are given. They are acquired by subtracting the background due to remagnetization, which is accounted for by a single exponential function, and by application of the Hamming window function.

\begin{figure}
		\includegraphics[width=0.8\columnwidth]{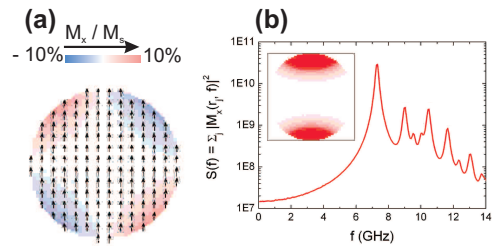}
	\caption{\label{fig4}(Color online) (a) Ground state of a 8 nm thick nickel nanodisc. (b) Frequency spectrum of this system and the end mode corresponding to the peak at lowest frequency.}
\end{figure}

\begin{figure}[h!]
		\includegraphics[width=0.7\columnwidth]{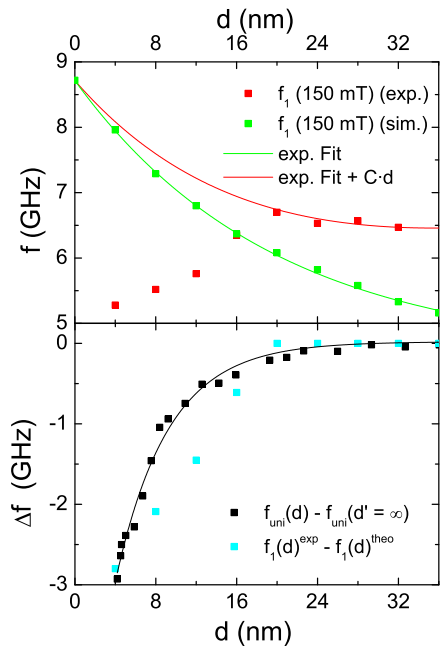}
	\caption{\label{fig5}(Color online) Thickness dependence of the dominant mode from experiment and simulation; while the simulated data follows an exponential function, the experimental data has to be corrected by a term linear in $d$ to account for dipole interaction (upper graph). At lower thicknesses, a surface or interface effect reduces the precessional frequencies as it is also found in a continuous film (lower graph); the line is a guide for the eyes only.}
\end{figure}

To understand which magnetic modes are excited we carry out micromagnetic simulations using the OOMMF code.\cite{donahue1999} A single nanodisc is modeled by means of a mesh of cubic cells with a size of $3\times 3\times 2(4)$ nm$^3$  for thicknesses below (above) 20 nm. The damping constant is chosen to be quite small ($\alpha=0.01$). The dynamics are induced by starting the simulation from an initial state that is slightly and statistically perturbed from the groundstate which is shown  in Fig. \ref{fig4} (a). The discrete solutions of the Landau Lifshitz Gilbert equation $\mathbf{M}(\mathbf{r}_j, t_i)$ delivered by OOMMF are Fourier transformed for each cell so that the excited modes can be identified. This way, the lowest frequency mode which happens to be the dominant mode is identified to be of end mode type, as it is depicted in Fig. \ref{fig4} (b) for an 8 nm thick structure and an external field of 150 mT.   It is assured that these results are not significantly altered in a simulation with vanishing out-of-plane component of the external field; therewith, the findings agree with the work of Zivieri and Stamps.\cite{zivieri:144422}

To compare this data with the experimental data, we determine the thickness dependence of this end mode and also extract the frequency of the broad maximum from the spectra found in the experiment; both are plotted versus nickel thickness in the upper graph of Fig. \ref{fig5}.  It turns out that the thickness dependence from the micromagnetic simulations follows an exponential function that coincides with the uniform precession frequency of a thin film for vanishing nickel thickness. The experimentally determined frequency dependence looks quite differently. However, the dipole interaction between nickel nanodiscs has been completely neglected within the simulations. If the dipole interaction is incorporated as an additional external magnetic field ($H_{\mathrm{ext}}+C\cdot d$) the frequency will alter to first order about an expression linear in nickel thickness d:
\begin{equation}
f\rightarrow f+\frac{\gamma_{0}}{2\pi}C\cdot d.
\end{equation}
By using the exponential fit to the simulated frequencies as expression for $f$,  the frequency behavior is adequately explained for thicknesses above 20 nm where the constant is determined to $\mu_0C=1.25(5)$ mT/nm. Therewith, we do not only recognize the experimentally dominant mode as an end mode, but also find an quantitative expression for describing the dipole interaction in this nanostructured film. The frequencies below 20 nm strongly deviate from this behavior. Here, a surface or interface effect comes into play that also reduces the frequency of the uniform precession of a continuous film prepared in the same manner. In the lower graph of Fig. \ref{fig5}, the frequency deviation, found in a continuous nickel wedge at lower thicknesses, is plotted together with the difference between the experimentally determined frequencies in the nanostructured wedge and the fitting curve in the upper graph. Considering the fact, that the demagnetizing factors for a disc are different from the ones of a continuous film, the similarity between the two plots is satisfying for relating both deviations to the same effect.

In conclusion, we have investigated the influence of the nanoscale on Laser-induced magnetization dynamics in nickel, both, on the time-scale of hundreds of femtoseconds as well as in the GHz regime. The fact that there seems to be no influence on the short time-scale might help to narrow possible microscopic models for the ultrafast demagnetization. On the other hand, the GHz dynamics of magnetization are significantly altered due to the nanostructuring. The thickness dependence of the excited magnetic mode allows us to quantitatively study the dipole interaction between single nickel nanodiscs. A possible coupling between magnetic  and elastic modes could not be proven or ruled out in this work so far.

Financial support by the Deutsche Forschungsgemeinschaft within the priority program SPP 1133 is gratefully acknowledged.


\end{document}